%Paper: hep-ph/9302261
%From: <malaney@chipmunk.cita.utoronto.ca>
%Date: Mon, 15 Feb 93 11:15:43 EST
%Date (revised): Thu, 18 Feb 93 13:10:27 EST

%\input figmac
%\input defs.tex
%\magnification \magstep1
\nopagenumbers
\hoffset=0 true in
\hsize=6.5 true in
\voffset=0 true in
\vsize=9 true in
\overfullrule=0pt
\abovedisplayskip=15pt
\belowdisplayskip=15pt
\abovedisplayshortskip=10pt
\belowdisplayshortskip=10pt
\baselineskip=20pt

\def\mh{m_H}

\def\mb{m_B}
\def\mf{m_F}

\def \Eh {E_H}

\def \Eb {E_B}
\def \Ef {E_F}
\def \nh {n_H}

\def \nb {n_B}
\def \nf {n_F}

\def \ph {p_H}

\def \pb {p_B}

\def \fh {f_H}

\def \fb {f_B}
\def \ff {f_F}

\def \nuh {\nu_H}

\def \Bbar {\overline{\rm B}}

\def \lase {{\rm lase}}

\def\gapp{\mathrel{\raise.3ex\hbox{$>$}\mkern-14mu
              \lower0.6ex\hbox{$\sim$}}}
\def\hbar{{\mathchar'26\kern-.5em{\it h}}}

% This puts pagenumbers at top of page

% This is for reference numbers

%gives greater than or approx
\def\gtorder{\mathrel{\raise.3ex\hbox{$>$}\mkern-14mu
             \lower0.6ex\hbox{$\sim$}}}
%give less than or approx
\def\ltorder{\mathrel{\raise.3ex\hbox{$<$}\mkern-14mu
             \lower0.6ex\hbox{$\sim$}}}

%\twelvepoint		%(NK) I don't have this
%\baselineskip=14truept plus 1pt minus 2pt
$  $

\vskip 0.5truein
\centerline {\bf Neutrino-Lasing in The Early Universe}
\noindent
\centerline{N. Kaiser$^{*}$, R. A. Malaney, G.D. Starkman}
\centerline{
CITA, University of Toronto, Toronto, ON, Canada M5S 1A7.}
\bigskip
\centerline{ABSTRACT}
Recently, Madsen has argued  that relativistic decays of
massive neutrinos into lighter fermions and bosons
may lead, via thermalization,
to the formation of a Bose condensate.  If correct, this could
generate mixed hot and cold dark matter, with
important consequences for structure formation.
 From a detailed study of such decays, we
arrive at substantially
different conclusions;  for a wide range of masses and decay times, we find
that
stimulated emission of bosons dominates the decay.
This phenomenon can best be described as a neutrino laser,  pumped by the QCD
phase transition.
We discuss the implications for structure formation
 and the dark-matter problem.
\vskip 0.2 true cm

Recent studies [1,2] suggest that a
mixed dark-matter (MDM) universe could  more
readily account for the  large-scale power observed
in galactic surveys [3] and recent COBE measurements of the microwave
background
fluctuations [4].
These developments have stimulated attempts to provide a
natural physical mechanism to generate MDM.
We focus here on the novel and interesting idea of Madsen [5] that
if a relativistic ``heavy" neutrino, $\nuh$, decays
into a light fermion, F, and boson, B, then a significant fraction of the
bosons may form a
condensate.

Consider the processes:
$$
\nu_H\rightleftharpoons {\rm F} + {\rm  B} \ \
; \bar\nu_H\rightleftharpoons \bar{\rm F} + \bar{\rm B} \ \ \ ,
\eqno (1)
$$
where  $\mf$ and $\mb$
are  small enough  ($\ltorder 100$
eV) so as not to overclose the universe and $\mf, \mb \ll \mh$.
Madsen argued that if the decay processes become effective
at $T = T_d$ (when the time-dilated free decay time for a $\nuh$ of typical
momentum is
equal to the age of the universe)  and if
the $\nuh$ are still relativistic,  then the  $\nuh$, F and B
populations would come into thermal (but not chemical)
equilibrium.
Solving for the parameters of the thermal distributions
and
assuming that the abundances of B and F were zero at $T_d$,
he finds that,
for a wide range of initial conditions, there is a
Bose condensate.

While highly suggestive, this heuristic argument is somewhat unsatisfactory.
First, the process is not
complete.  In coming to the supposed equilibrium state only about half of the
$\nuh$'s decay.
What happens to the remaining $\nuh$'s is unclear.
Second, a sudden equilibration at $T_d$ may be  an inadequate
description. To address these issues we have solved the Boltzmann equations
for the reactions (1).

% We find that there is a region of
% parameter space for which  the equilibration envisaged by Madsen
% takes place, but this requires $T_d \gtorder E_* \simeq \mh^2 /2\mb$, rather
%%than
% $T_d > \mh$. After they equilibrate, the populations remain stable until the
% remaining $\nuh$'s become non-relativistic and then decay into
% hot ($E \sim \mh/2$) bosons.
We find that if the decay rate is sufficiently high
($T_d > \sqrt{\mh \mb}$) then the equilibration will be preceded by a
burst of stimulated decays into very cold ($p \ll T$) bosons.
This is an exponentially unstable process
{\it i.e.}\ a {\it neutrino-laser}.
As a result of this non-thermal process,
close to half of the $\nuh$'s decay.
The  resulting $\nuh$ and F distributions have a
`grey-body' form and the cold B's are also non-thermal,
with an abundance well in excess of the thermal-equilibrium
Bose-condensate.

Allowing only the reactions (1),
the  evolution of the occupation number  distributions $f_i$,  $i=$H,F,B,
are described by the following Boltzmann equations:
$$
{\dot f_i} = ( h_i\mh^2 \Gamma_0 / m_0 E_i p_i) \int
dE_{j\neq{i}} [\fh (1 - \ff)(1+ \fb) - \fb \ff(1 - \fh)]  \eqno(2)
$$
where ${\dot f_i}$ is the time derivative of $f_i$
at a fixed comoving momentum,
$h_T = -1$, $h_F = 1$ and $h_B = 2 (1)$ if $B =(\neq) \Bbar$, $p_i$ is the
physical
three-momentum, and $m_0/2$ is the three-momentum of the
decay products in the $\nuh$
rest-frame with $ m_0^2 = \mh^2 - 2(m_B^2 + m_F^2) + (m_B^2 - m_F^2)^2/\mh^2 $.
$\Gamma_0$ is the free decay rate for a $\nuh$ at rest.
The integration is over the energy-conserving plane $\Eh = \Ef + \Eb$, with
limits
on $\Eb,\Ef$
$$
E_k^{\pm} = (m_0 / 2 \mh)
(\Eh \sqrt{1 + 4 m_k^2 / m_0^2} \pm \ph)
\eqno(3).
$$
($k=B,F$)  as shown in figure 1.
These limits follow from purely kinematic considerations.
For high $\Eh$, F and B both
come off in the forward direction with similar energies. For
$$\Eh = E_* = (\mh^2 + \mb^2 - \mf^2) / 2\mb \simeq \mh^2 /2\mb$$
zero-momentum bosons are accessible.
If the final states were empty the
$\nuh$'s would decay into products with energies distributed uniformly
on the range (3). Equation (2) then follows straightforwardly from the
inclusion of the
quantum mechanical statistical weights for the forward and inverse
decay processes.

To set the
initial conditions
we assume that at some early time $\nuh$, F and B were all in
chemical and thermal equilibrium, but that the F and B
decoupled prior to the QCD phase transition
at $T\approx 100$ MeV while the $\nuh$ did not;
this specifically requires ${\rm F} \neq \nu_e,\nu_\mu$.
Following the phase transition
 the temperature of the $\nuh$'s was increased  by a factor
$\eta^{1/3}\simeq 2$ relative to the F's and B's,
where $\eta$ is the ratio of statistical weights before and after the
phase transition.
 Equations (2) can be
 integrated numerically, but we
can gain insight into the general properties of the solution by
considering some limiting regimes.

Consider the equation for $\dot \fh$.
Since $1 - \ff \le 1$ and the range of the integral is
$\Eb^{+} - \Eb^{-} = m_0 \ph / \mh$, the
rate for terms independent of $\fb$ is $\sim \mh \Gamma_0 / \Eh$. This is just
the time dilated free decay rate for a typical energy $\nuh$ and therefore
will be negligible for $T \gg T_d \equiv \min(T_0, (T_0^2 \mh)^{1/3})$, where
$T_0$ is such that $H(T_0) = \Gamma_0$.
The terms involving $\fb$, however, can be important long before $T_d$.
Neglecting the other terms we have
$$
\dot \fb = (h_B \Gamma_0 \mh^2 \fb/ m_0 \Eb \pb)
\int dE (\fh  - \ff )
\eqno (4)
$$
so $\fb$ will grow exponentially, driven by any initial imbalance
between the $\nuh$'s and F's.  The limits on the integral in (4) are
$\Eh^{\pm}(\Eb)$, which are given implicitly by Eq~(3).
The prefactor $1/\Eb\pb$ suggests the growth rate will be greatest for
the lowest momentum bosons which are accessible.  Consider first
the case $T \gg E_*$ which must hold at sufficiently early times.
The minimum boson energy for a typical energy $\nuh$ ($\Eh \sim T$) is
$\Eb^- \sim (\mb/\mh)^2 \Eh$.  Setting $\Eb,\pb \sim \Eb^-$ and since the
integral in
equation (4) will be on the order $\Eh \sim T$ we obtain
$$
\dot \fb
\simeq (\Gamma_0 \mh^5 / \mb^4 T) \fb
\eqno (5)
$$
so the growth rate is $\Gamma_\lase \simeq \Gamma_0 (\mh/\mb)^3 (E_* / T)$.
This increases with time while the expansion rate decreases, so these will be
equal at
$$
T_\lase \simeq (\mh^5 T_0^2 / \mb^4)^{1/3}
\eqno (6)
$$
We have assumed here that
$T \gg E_*$.  For $T \ll E_*$ we find that the growth rate decreases faster
than the Hubble rate so if the process is not effective at $T \simeq E_*$ then
it never will be.
The condition  $T_\lase \gtorder E_*$ sets a lower limit on the decay rate, or
equivalently on the decay temperature: $T_d > \sqrt{\mh \mb}$.

What we have here is a  neutrino laser;
the low momentum boson occupation numbers will grow exponentially via
stimulated decays,
feeding off any initial imbalance between the $\nuh$'s and F's, and terminating
when this is driven to zero.  The momentum of the bosons produced is
$\pb \simeq (\mb^2/\mh^2) T$, so for a large mass ratio these will be very cold
compared to the typical thermal energy.

The
lasing will  result in
$\fh =\ff = (\fh^o +\ff^o) / 2$  (superscript indicates initial values)
which is a fermion analogue of a `grey body' spectrum,
and the total number of
cold bosons produced is just $({\nh}^o - {\nf}^o) / 2 = (1 - \eta^{-1})
{\nh}^o / 2$.
The number of hot bosons is just the initial number ${\nb}^o = 4 \nh^o
/ (3 \eta h_B)$. This then gives
 the fraction of cold bosons after lasing to be $\simeq 80(65)$\% for
$B=(\ne)\Bbar$.
The result of lasing is therefore qualitatively similar to the
thermalization calculation of Madsen, but with important differences:
The lasing process occurs much earlier than thermalization; the initial
hot bosons are unaffected and the decays occur exclusively into the cold
component.
In the equilibration calculation, only about half of the decays go into
the cold phase, and the cold fraction is
$\simeq 42(26)\%$,
about half the yield from lasing (these numbers are slightly
different from those calculated by Madsen as we have allowed for the
finite initial boson abundance).

After lasing the temperature will eventually fall to $T_d$.  What happens then
depends $T_d$. For $T_d < E_*$
the low momentum bosons produced by lasing will be effectively decoupled (they
can
only be reached from $\Eh \sim E_* \gg T$ and so this coupling will be
exponentially
suppressed). The hot bosons and fermions will equilibrate, but they do this
without
generating any further bosons.
For $T_d > E_*$ on the other hand, the laser generated bosons are still
accessible
from typical energy decays, so we
expect that some of the cold bosons will be reabsorbed and that the post-$T_d$
distributions will adopt the form predicted by Madsen.

In either case the decay process is not yet finished since there is still
a finite $\nuh$ abundance.
The remaining $\nuh$'s (58\% of the original abundance)
decay into hot B's and F's once $T \ltorder \min(\mh,T_d)$.
For lasing, the final fraction of
cold bosons is
$(1-\eta^{-1}) / (2 + 8 / 3 \eta h_B) \simeq$ 37(34)\%.
For equilibration ($T_d > E_*$) the numbers are 19(13)\%.
When and how the final decays occur depends on $T_d$.  For $T_d > E_*$,
the populations equilibrate at $T_d$ and are then stable until $T \sim \mh$
when the heavy neutrinos go non-relativistic at which time they decay.
For $\mh < T_d < E_*$ we expect the hot particles to come to thermal
equilibrium at $T \sim T_d$ without generating any further
cold bosons.  The remaining $\nuh$'s decay at $T\sim \mh$,
with no subsequent equilibration.
For $\sqrt{\mh \mb} < T_d < \mh$ no equilibration occurs and the particles
decay at $T_d$ when they are non-relativistic.
The last possibility is $T_d < \sqrt{\mh \mb}$ in which case no cold bosons are
produced and the $\nuh$'s decay into very hot products at $T_d$.

These predictions are confirmed by our numerical solution of (2).
In Figure 2 the evolution of the occupation numbers
are plotted for masses and decay temperature chosen to illustrate the lasing
phenomenon.

In all of this one must be
wary that  the new fermionic
degree of freedom F does not take up its full statistical weight
prior to the weak interaction freeze-out,
 thereby
violating the bounds from standard big bang nucleosynthesis [6].
(The B's are not of concern if the stimulated decay is dominant
since they are very low energy.)
We therefore require $T_\lase <$ 2.3 MeV since
the number changing reactions for the neutrinos
are decoupled below this temperature. From equation (6) and the definition of
$T_d$ this requires $T_d < (2.3{\rm MeV})^{3/2} \mb^2 \mh^{-5/2}$ if
$T_d < \mh$ and $T_d < 2.3{\rm MeV}(\mb/\mh)^{4/3}$ otherwise.
The former, together with the lower bound $T_d > \sqrt{\mh\mb}$
sets an upper limit
$\mh \ltorder \sqrt{2 {\rm MeV} \mb}
\sim 10$keV for a boson mass which would close the universe.
These, together with the  requirement that
$\mh>m_B, m_F$,
are shown in  Fig.\ 3, which displays the region of $T_d-\mh$
space where cold bosons are produced.

The fraction of cold particles is important
for structure formation,
and provides a firm (modulo
factors such as $h_B$, $\eta$) prediction of the lasing model.
Also important, however, is $m_B$ and the momentum distribution of the
hot bosons, as they will determine the maximum Jeans mass[7].
The attraction of the
70/30 cold/hot ratio MDM model [1] is mostly that the low hot-fraction
 reduces the mass of the neutrino by a factor $\simeq$3,  therefore
 increasing
the maximum Jeans mass by a factor of $\simeq$10.  Here the final abundance
of bosons is just $(1 + 4  / 3  \eta h_B)\nh^o$.
If the bosons alone
constitute the dark matter then their mass will be only slightly less than the
standard value.  However, the boson momentum distribution will also differ from
the
standard value.  For $T_d > E_*$ the mean momentum of the hot bosons is
almost identical to the standard value so the maximum Jeans mass
will be very similar to that in the standard HDM model.
This may be problematic.
In the case of lasing with $T_d > \mh$ the hot bosons are $\simeq$ 40\%
{\it cooler}
than in standard HDM making matters even worse.  However, for
the lower triangle $\sqrt{\mh \mb} < T_d < \mh$,
the temperature of the hot bosons can be larger
than the ambient temperature by a factor $\sim \mh / T_d$.  This can be as
large as $\sim \sqrt{\mh / \mb} \ltorder (2 {\rm MeV} /
\mb)^{1/4} \simeq 20$. It is also possible that the fermion and boson masses
are similar, or that other particles,  {\it e.g.}\ the $\nu_\mu$ decay into
the same products; in either case this would increase the
maximum Jeans mass.

We note that
a number of variations on the preceding
discussion exist,  which may have
different quantitative predictions.  For example, if $\nuh=\nu_\mu$,
and $\nu_\mu$ is
heavier than $\nu_\tau$, then the QCD phase transition would pump
the $\nu_\mu-\nuh$ system, because charged-current production of
$\nu_\tau$ is energetically disfavored;
 or the $\nuh$, F and B could all be new particles outside
the standard model.
Investigation of such possibilities is beyond the scope of the present work.

We have thus far only
 considered  the scenario where F$\ne\nu_e,\nu_\mu$.  Consider  briefly
the scenario where
the light fermion is one of the known neutrinos, and is therefore coupled
at $T_{\rm QCD}$. Madsen's calculations [5] indicate that in this scenario too,
 Bose condensation may occur. With regard to our calculations note that
 there is no initial fermion imbalance to drive the
lasing.  In fact, we conclude that
direct solution of the Boltzmann equations (Eq.~2)  results in no
condensation.

In conclusion:
we have shown that for decay temperatures $T_d > \sqrt{\mh \mb}$
that thermalization is inevitably preceded by a runaway stimulated
decay into cold ($p \sim (\mb/\mh)^2 T$) bosons.
This `neutrino lasing' provides a  relatively natural way to generate MDM.
For decay temperatures $T_d \gg \mh^2 / \mb$ the populations will
thermally equilibrate, some of the cold bosons will be reabsorbed, and the
final
cold boson fraction is $\simeq15\%$.  For
$T_d \ll \mh^2 / \mb$ the laser generated bosons survive to the
present and the cold fraction is $\simeq35\%$.
If the boson alone constitutes the dark matter,
then thermalization results in a maximum Jeans mass very
similar to that in the standard HDM model; in contrast, the lasing scenario
allows the interesting possibility of a higher Jeans mass.

We greatly appreciate numerous discussions with Arif Babul, Dick Bond, Malcolm
Butler,
Savas Dimopoulos, Lev Kofman, Lenny Susskind, Chris Thomson, Scott Tremaine and
 Ned Wright.
\medskip
\noindent
REFERENCES
\medskip
\noindent
$^*$CIAR Cosmology Program
\medskip \noindent
[1] M. Davis, F. J. Summers and D. Schlegel, Nature {\bf 359}, 393 (1992).
\medskip
\noindent
[2] A. N. Taylor and M. Rowan-Robinson, Nature {\bf 359}, 396 (1992).
\medskip
\noindent
[3] S. J. Maddox, G. P. E. Efstathiou, W. Sutherland and J. Loveday,
Mon. Not. R. Astr. Soc., {\bf 242}, p43 (1990).
\medskip
\noindent
[4] G. Smoot {\it et al.}, Astrophys. J., {\bf 396}, L1 (1992);
E. L. Wright {\it et al.}, Astrophys. J., {\bf 396}, L13 (1992)
\medskip
\noindent
[5] J. Madsen, Phys. Rev. Lett. {\bf 69}, 571 (1992).
\medskip
\noindent
[6] M. Smith, L. Kawano and R. A. Malaney, Astrophys. J. Supp. May 1 issue
(1993),
(and refs. therein).
\medskip
\noindent
[7] J. R. Bond, G. P. Efstathiou and J. Silk,
Phys. Rev. Lett. {\bf 45}, 1980 (1980).
\vfill
\eject

\centerline{{\bf FIGURE CAPTIONS}}

\noindent
1. Allowed $\Eb$ for decay of $\nuh$ with
energy $\Eh$ (logarithmic scales). Of particular importance for
our analysis is the form of $\Eb^-$ since the lasing occurs into
the lowest momentum states which are accessible. Low energy $\nuh$'s
decay into high energy products.
Particles with $\Eh = E_* \simeq \mh^2 / \mb$ can decay into
zero momentum bosons and for $\Eh > E_*$, $\Eb^- \simeq (\mb/\mh)^2 \Eh$.
\hfil\break

\noindent
2. Upper panels show $p^3 f$, the number of
particles per $\log(p)$ for three output times: the initial time, after lasing
and after all $\nuh$'s have decayed.  The masses
were $\mh;\ \mf;\ \mb = 1;\ 0.03;\ 0.03$ and $T_d = 1$.
The lower panel shows the total number of bosons as a function of time.
The lasing epoch, which injects cold bosons exclusively, followed by
a trans-relativistic decay epoch, injecting hot bosons, are evident.
Most striking is the bimodality
of the boson distribution which is clearly divided into  hot and cold
components.
\hfil\break

\noindent
3.  Allowed regions for generating mixed dark
matter (logarithmic scales). In the upper triangle lasing occurs, but is then
followed by equilibration resulting in a reduction in the number of cold
bosons.
In the lower triangles lasing occurs.
Above the dashed line, the sequence of events is lasing;
equilibration of the hot components; and finally decays at $T\sim \mh$.
Below the dashed line,
the particles become non-relativistic before the decay process terminates.

\vfil \eject \bye